\documentclass[twocolumn,english,prb,aps,fleqn,eqsecnum,amsmath,amssymb,floats,secnumarabic,floatfix]{revtex4-1}
\usepackage[T1]{fontenc}
\usepackage[latin9]{inputenc}
\usepackage{color}
\usepackage{babel}
\usepackage{units}
\usepackage{amsmath}
\usepackage{graphicx}
\usepackage[unicode=true,
 bookmarks=true,bookmarksnumbered=false,bookmarksopen=false,
 breaklinks=true,pdfborder={0 0 0},backref=false,colorlinks=true]
 {hyperref}
\hypersetup{pdftitle={The Hubbard model beyond the two-pole approximation: a Composite Operator Method study},
 pdfauthor={Adolfo Avella}}

\makeatletter

\providecommand{\tabularnewline}{\\}

\@ifundefined{textcolor}{}
{%
 \definecolor{BLACK}{gray}{0}
 \definecolor{WHITE}{gray}{1}
 \definecolor{RED}{rgb}{1,0,0}
 \definecolor{GREEN}{rgb}{0,1,0}
 \definecolor{BLUE}{rgb}{0,0,1}
 \definecolor{CYAN}{cmyk}{1,0,0,0}
 \definecolor{MAGENTA}{cmyk}{0,1,0,0}
 \definecolor{YELLOW}{cmyk}{0,0,1,0}
}

\usepackage{url}

\makeatother

\begin{document}

\title{COM(3p) solution of the 2D Hubbard model: momentum resolved quantities}

\author{Adolfo Avella}

\affiliation{Dipartimento di Fisica ``E.R. Caianiello'', Universit� degli Studi
di Salerno, I-84084 Fisciano (SA), Italy}

\affiliation{Unit� CNISM di Salerno, Universit� degli Studi di Salerno, I-84084
Fisciano (SA), Italy}

\affiliation{CNR-SPIN, UoS di Salerno, I-84084 Fisciano (SA), Italy}
\begin{abstract}
Recently, within the framework of the Composite Operator Method, it
has been proposed a three-pole solution for the two-dimensional Hubbard
model \cite{Avella_14}, which is still considered one of the best
candidate model to microscopically describe high-$T_{c}$ cuprate
superconductors. The operatorial basis comprise the two Hubbard operators
(complete fermionic local basis) and the electronic operator dressed
by the nearest-neighbor spin fluctuations. The effectiveness of the
approximate solution has been proved through a positive comparison
with different numerical methods for various quantities. In this article,
after recollecting the main analytical expressions defining the solution
and the behavior of basic local quantities (double occupancy and chemical
potential) and of the quasi-particle energy dispersions, we resolve
and analyze the momentum components of relevant quantities: filling
(i.e. the momentum distribution function), double occupancy and nearest-neighbor
spin correlation function. The analysis is extended to COM(2p) solutions
that will be used as primary reference. Thanks to this, the role played
by the third field, with respect to the two Hubbard ones, in determining
the behavior of many relevant quantities and in allowing the extremely
good comparison with numerical results is better understood giving
a guideline to further improve and, possibly, optimize the application
of the COM to the Hubbard model.
\end{abstract}
\maketitle

\section{Introduction}

The cuprate high-$T_{c}$ superconductors \cite{Bednorz_86} still
lack a widely accepted and unifying microscopic description of their
anomalous behavior experimentally observed, mainly in the underdoped
region, in almost all experimentally measurable physical properties
\cite{Timusk_99,Basov_99,Orenstein_00,Damascelli_03,Shen_05,Eschrig_06,Kanigel_06,Lee_06,Valla_06,Doiron-Leyraud_07,LeBoeuf_07,Hossain_08,Sebastian_08,Meng_09,Laliberte_11,Ramshaw_11,Riggs_11,Sebastian_11a,Sebastian_12a}.
Non-Fermi-liquid response, quantum criticality, pseudogap formation,
ill-defined Fermi surface, kinks in the electronic dispersion, $\ldots$
cannot be explained by standard many-body theory within the Fermi-liquid
framework by means of diagrammatic expansions and remain controversially
debated \cite{Lee_06,Tremblay_06,Sebastian_12a}. Strong electronic
correlations, competition between localization and itinerancy, Mott
physics, and low-energy spin excitations are considered key ingredients
necessary to explain these anomalous features and the Hubbard model
\cite{Hubbard_63} contain all of them by construction.

The Hubbard model\cite{Hubbard_63} together with its relevance to
real materials, in particular cuprate high-$T_{c}$ superconductors,
has always raised a more fundamental and theoretical interest as it
is universally considered the prototypical model for strongly correlated
systems. Unfortunately, although many trials have been made, no analytical
approximation method can be considered to have given a clear and definitive
answer to the very many relevant issues raised by this very simple
model. Numerical approaches \cite{Avella_13a} are fundamental for
benchmarking and fine tuning analytical theories and for establishing
which are those capable to deal with the quite complex phenomenology
of the Hubbard model. Unfortunately, numerical techniques cannot explore,
because of their limited resolution in frequency and momentum, the
most relevant regime of model parameters (small doping, low temperature
and large on-site Coulomb repulsion) where one expects strong electronic
correlations to dominate the physics of the system. As regards analytical
and semi-analytical (i.e. embedding a numerical core) theories \cite{Avella_12},
a few are definitely worth mentioning: the work of Mori \cite{Mori_65},
Hubbard \cite{Hubbard_63,Hubbard_64,Hubbard_64a}, Rowe \cite{Rowe_68},
Roth \cite{Roth_69}, Tserkovnikov \cite{Tserkovnikov_81,Tserkovnikov_81a},
the Gutzwiller approximation \cite{Gutzwiller_63,Gutzwiller_64,Gutzwiller_65,Brinkman_70},
the slave boson method \cite{Barnes_76,Coleman_84,Kotliar_86}, the
spectral density approach \cite{Kalashnikov_69,Nolting_72}, the two-particle
self-consistent approach \cite{Tremblay_06}, the RPA and equations-of-motion
based techniques \cite{Chubukov_04,Prelovsek_05,Plakida_06}, the
dynamical mean-field theory (DMFT) \cite{Metzner_89,Georges_92,Georges_96},
the DMFT$+\Sigma$ approach \cite{Sadovskii_05,Kuchinskii_05,Kuchinskii_06}
as well as all cluster-DMFT-like theories\cite{Maier_05} (the cellular-DMFT
\cite{Kotliar_01a}, the dynamical cluster approximation \cite{Hettler_98}
and the cluster perturbation theory \cite{Senechal_00}).

We have also been developing a systematic approach, the composite
operator method (COM) \cite{Theory,Avella_11a}, to study highly correlated
systems. In the last fifteen years, COM has been applied to several
models and materials: Hubbard \cite{Hub-redux,Odashima_05,Avella_14},
$p$-$d$ \cite{p-d}, $t$-$t'$-$U$ \cite{ttU-redux}, extended
Hubbard ($t$-$U$-$V$) \cite{tUV-redux}, Kondo \cite{Villani_00},
Anderson \cite{Anderson-redux}, two-orbital Hubbard \cite{Plekhanov_11},
Ising \cite{Ising-redux}, $J_{1}-J_{2}$ \cite{Avella_08a,J1J2-redux},
Cuprates \cite{Cuprates-NCA,Avella_07,Avella_07a,Avella_08,Avella_09},
etc The Composite Operator Method (COM) \cite{Theory,Avella_11a}
has the advantage to be completely microscopic, exclusively analytical,
and fully self-consistent. COM recipe uses three main ingredients
\cite{Theory,Avella_11a}: \emph{composite} operators\emph{, algebra}
constraints and \emph{residual} self-energy. Composite operators are
products of electronic operators and describe the new elementary excitations
appearing in the system owing to strong correlations. According to
the system under analysis \cite{Theory,Avella_11a}, one has to choose
a set of composite operators as operatorial basis and rewrite the
electronic operators and the electronic Green's function in terms
of this basis. Algebra Constraints are relations among correlation
functions dictated by the non-canonical operatorial algebra closed
by the chosen operatorial basis \cite{Theory,Avella_11a}. Other ways
to obtain algebra constraints rely on the symmetries enjoined by the
Hamiltonian under study, the Ward-Takahashi identities, the hydrodynamics,
etc \cite{Theory,Avella_11a}. Algebra Constraints are used to compute
unknown correlation functions appearing in the calculations. Interactions
among the elements of the chosen operatorial basis are described by
the residual self-energy, that is, the propagator of the residual
term of the current after this latter has been projected on the chosen
operatorial basis \cite{Theory,Avella_11a}. According to the physical
properties under analysis and the range of temperatures, dopings,
and interactions you want to explore, one has to choose an approximation
to compute the residual self-energy. In the last years, we have been
using the $n-$pole Approximation \cite{Hub-redux,Odashima_05,p-d,ttU-redux,tUV-redux,Plekhanov_11,Ising-redux,Cuprates-NCA,Avella_14},
the Asymptotic Field Approach \cite{Villani_00,Anderson-redux} and
the Non-Crossing Approximation (NCA) \cite{Avella_07,Avella_07a,Avella_08,Avella_09}.

In this article, we first recollect the main analytical expressions
defining the COM(3p) approximation for the 2D Hubbard model (Sec.\ \ref{sec:Theory}).
More details can be found in \cite{Avella_14}. Then, we set the stage
by reporting both (i) basic local quantities (Sec.\ \ref{sec:Results:D_mu}:
double occupancy and chemical potential), comparing them with numerical
and semi-analytical methods to asses the solution and characterize
it, and (ii) the quasi-particle dispersions (Sec.\ \ref{sec:Results:Ek}).
These latter, in particular, together with the comparison to COM(2p)
solutions, will allow to analyze and understand the behavior of the
momentum resolved components of relevant quantities: filling (i.e.
the momentum distribution function), double occupancy and nearest-neighbor
spin correlation function (Sec.\ \ref{sec:Results:nk_Dk_chiask}).
Finally, in Sec.\ \ref{sec:Summary}, we draw some conclusions.

\section{Theory\label{sec:Theory}}

\subsection{Hamiltonian}

The Hamiltonian of the two-dimensional Hubbard model reads as 
\begin{multline}
H=-4t\sum_{\mathbf{i}}c^{\dagger}\left(i\right)\cdot c^{\alpha}\left(i\right)\\
+U\sum_{\mathbf{i}}n_{\uparrow}\left(i\right)n_{\downarrow}\left(i\right)-\mu\sum_{\mathbf{i}}n\left(i\right)\label{eq:Ham}
\end{multline}
where $c^{\dagger}\left(i\right)=\begin{pmatrix}c_{\uparrow}^{\dagger}\left(i\right) & c_{\downarrow}^{\dagger}\left(i\right)\end{pmatrix}$
is the electronic field operator in spinorial notation and Heisenberg
picture ($i=\left(\mathbf{i},t_{i}\right)$). $\cdot$ and $\otimes$
stand for the inner (scalar) and the outer products, respectively,
in spin space. $\mathbf{i}=\mathbf{R_{i}}=\left(i_{x},i_{y}\right)$
is a vector of the two-dimensional square Bravais lattice, $n_{\sigma}\left(i\right)=c_{\sigma}^{\dagger}\left(i\right)c_{\sigma}\left(i\right)$
is the particle density operator for spin $\sigma$ at site $\mathbf{i}$,
$n\left(i\right)=\sum_{\sigma}n_{\sigma}\left(i\right)=c^{\dagger}\left(i\right)\cdot c\left(i\right)$
is the total particle density operator at site $\mathbf{i}$, $\mu$
is the chemical potential, $t$ is the hopping integral and the energy
unit hereafter, $U$ is the Coulomb on-site repulsion and $\alpha_{\mathbf{ij}}$
is the projector on the nearest-neighbor sites 
\begin{align}
\alpha_{\mathbf{ij}} & =\frac{1}{N}\sum_{\mathbf{k}}\mathrm{e}^{\mathrm{i}\mathbf{k}\cdot(\mathbf{R_{i}}-\mathbf{R_{j}})}\alpha\left(\mathbf{k}\right)\label{eq:alpha-ij}\\
\alpha\left(\mathbf{k}\right) & =\frac{1}{2}\left[\cos\left(k_{x}a\right)+\cos\left(k_{y}a\right)\right]\label{eq:alpha-k}
\end{align}
where $\mathbf{k}$ runs over the first Brillouin zone, $N$ is the
number of lattice sites and $a$ is the lattice constant, which will
be set to one for the sake of simplicity. For any operator $\Phi\left(i\right)$,
we use the notation $\Phi^{\kappa}\left(i\right)=\sum_{\mathbf{j}}\kappa_{\mathbf{ij}}\Phi\left(\mathbf{j},t_{i}\right)$
where $\kappa_{\mathbf{ij}}$ can be any function of the two sites
$\mathbf{i}$ and $\mathbf{j}$ and, in particular, a projector over
the cubic harmonics of the lattice: e.g. $c^{\alpha}\left(i\right)=\sum_{\mathbf{j}}\alpha_{\mathbf{ij}}c\left(\mathbf{j},t_{i}\right)$.

\subsection{Basis and equations of motion}

According to COM prescription \cite{Theory,Avella_11a}, we have chosen
as composite basic field 
\begin{equation}
\psi\left(i\right)=\begin{pmatrix}\psi_{1}\left(i\right)\\
\psi_{2}\left(i\right)\\
\psi_{3}\left(i\right)
\end{pmatrix}=\begin{pmatrix}\xi\left(i\right)\\
\eta\left(i\right)\\
c_{s}\left(i\right)
\end{pmatrix}\label{eq:basis}
\end{equation}
where $\eta\left(i\right)=n\left(i\right)c\left(i\right)$ and $\xi\left(i\right)=c\left(i\right)-\eta\left(i\right)$
are the Hubbard operators and $c_{s}\left(i\right)=n_{k}\left(i\right)\sigma_{k}\cdot c^{\alpha}\left(i\right)$
is the electronic operator dressed by the nearest-neighbor spin fluctuations,
which are expected to be the most relevant fluctuations, compared
to charge and pair ones, in determining the fundamental response and
the important features of the system under analysis \cite{Avella_14}.
This assumption has been proved to be definitely valid \cite{Avella_14}
in the parameter regime where the electronic correlations are expected
to be very strong: large $U$, small doping $\delta=1-n$ and low
temperature $T$. In absence of correlations, or for the very weak
ones, no type of fluctuations is relevant. $n_{\mu}\left(i\right)=c^{\dagger}\left(i\right)\cdot\sigma_{\mu}\cdot c\left(i\right)$
is the charge- ($\mu=0$) and spin- ($\mu=1,2,3=k$) density operator,
$\sigma_{\mu}=\left(1,\vec{\sigma}\right)$, $\sigma^{\mu}=\left(-1,\vec{\sigma}\right)$,
$\sigma_{k}$ with $\left(k=1,2,3\right)$ are the Pauli matrices. 

The field $\psi(i)$ satisfies the following equation of motion 
\begin{equation}
\mathrm{i}\frac{\partial}{\partial t}\psi\left(i\right)=\begin{pmatrix}-\mu\xi\left(i\right)-4tc^{\alpha}\left(i\right)-4t\pi\left(i\right)\\
\left(U-\mu\right)\eta\left(i\right)+4t\pi\left(i\right)\\
-\mu c_{s}\left(i\right)+4t\kappa_{s}\left(i\right)+U\eta_{s}\left(i\right)
\end{pmatrix}
\end{equation}
where the higher-order composite fields $\pi\left(i\right)$, $\kappa_{s}\left(i\right)$
and $\eta_{s}\left(i\right)$ are defined as 
\begin{align}
 & \pi\left(i\right)=\frac{1}{2}n_{\mu}(i)\sigma^{\mu}\cdot c^{\alpha}\left(i\right)+c^{\dagger\alpha}\left(i\right)\cdot c\left(i\right)\otimes c\left(i\right)\\
 & \kappa_{s}\left(i\right)=-n_{k}\left(i\right)\sigma_{k}\cdot c^{\alpha^{2}}\left(i\right)\nonumber \\
 & +\left(c^{\alpha\dagger}\left(i\right)\cdot\sigma_{k}\cdot c\left(i\right)-c^{\dagger}\left(i\right)\cdot\sigma_{k}\cdot c^{\alpha}\left(i\right)\right)\sigma_{k}\cdot c^{\alpha}\left(i\right)\\
 & \eta_{s}\left(i\right)=n_{k}\left(i\right)\sigma_{k}\cdot\eta^{\alpha}\left(i\right)
\end{align}
It is clear now that $c_{s}\left(i\right)$ has been chosen proportional
to the \emph{spin} component of $\pi\left(i\right)$. Accordingly,
we define $\bar{\pi}\left(i\right)=\pi\left(i\right)-\frac{1}{2}c_{s}\left(i\right)$.

\subsection{Current projection (pole approximation)}

The current $J\left(i\right)=\mathrm{i}\frac{\partial}{\partial t}\psi\left(i\right)=\left[\psi\left(i\right),H\right]$
of the basis $\psi(i)$ can be approximated 
\begin{equation}
J\left(i\right)\cong\sum_{\mathbf{j}}\varepsilon\left(\mathbf{i},\mathbf{j}\right)\psi\left(\mathbf{j},t\right)\label{eq:J}
\end{equation}
projecting the current $J\left(i\right)$ on the basis $\psi\left(i\right)$.
$\varepsilon\left(\mathbf{i},\mathbf{j}\right)$ is named energy matrix
and can be computed by means of the equation 
\begin{multline}
\left\langle \left\{ J\left(\mathbf{i},t\right),\psi^{\dagger}\left(\mathbf{j},t\right)\right\} \right\rangle \\
=\sum_{\mathbf{j}}\varepsilon\left(\mathbf{i},\mathbf{j}\right)\left\langle \left\{ \psi\left(\mathbf{i},t\right),\psi^{\dagger}\left(\mathbf{j},t\right)\right\} \right\rangle \label{eq:dJ}
\end{multline}
where $\left\langle \cdots\right\rangle $ stands for the thermal
average taken in the grand-canonical ensemble: 
\begin{equation}
\varepsilon(\mathbf{k})=m(\mathbf{k})I^{-1}(\mathbf{k})\label{eq:m-I}
\end{equation}
where
\begin{align}
\varepsilon\left(\mathbf{i},\mathbf{j}\right) & =\frac{1}{N}\sum_{\mathbf{k}}\mathrm{e}^{\mathrm{i}\mathbf{k}\cdot(\mathbf{R_{i}-R_{j}})}\varepsilon\left(\mathbf{k}\right)\\
I\left(\mathbf{i},\mathbf{j}\right) & =\left\langle \left\{ \psi\left(\mathbf{i},t\right),\psi^{\dagger}\left(\mathbf{j},t\right)\right\} \right\rangle \nonumber \\
 & =\frac{1}{N}\sum_{\mathbf{k}}\mathrm{e}^{\mathrm{i}\mathbf{k}\cdot(\mathbf{R_{i}-R_{j}})}I\left(\mathbf{k}\right)\\
m\left(\mathbf{i},\mathbf{j}\right) & =\left\langle \left\{ J\left(\mathbf{i},t\right),\psi^{\dagger}\left(\mathbf{j},t\right)\right\} \right\rangle \nonumber \\
 & =\frac{1}{N}\sum_{\mathbf{k}}\mathrm{e}^{\mathrm{i}\mathbf{k}\cdot(\mathbf{R_{i}-R_{j}})}m\left(\mathbf{k}\right)
\end{align}
 Since $\psi\left(i\right)$ is made up of composite operators, the
normalization matrix $I\left(\mathbf{k}\right)$ is not the identity
matrix as it happens for the original electronic field operator. Hereafter,
we will use the very convenient notation $I_{\phi\varphi}\left(\mathbf{i},\mathbf{j}\right)=\left\langle \left\{ \phi\left(\mathbf{i},t\right),\varphi^{\dagger}\left(\mathbf{j},t\right)\right\} \right\rangle $,
which generalizes the definition of the normalization matrix ($I\left(\mathbf{i},\mathbf{j}\right)=I_{\psi\psi}\left(\mathbf{i},\mathbf{j}\right)$)
and of the $m$-matrix ($m\left(\mathbf{i},\mathbf{j}\right)=I_{J\psi}\left(\mathbf{i},\mathbf{j}\right)$)
and provide the operator space of a scalar product.

\subsection{Green's and correlation functions}

By using the projection of the source (\ref{eq:J}), that is, by working
in the framework of a three-pole approximation, and by introducing
the Fourier transform $\mathcal{F}_{\mathbf{k}\omega}\left[\cdots\right]$,
the retarded thermodynamic Green's functions 
\begin{multline}
G\left(i,j\right)=\left\langle \mathcal{R}\left[\psi\left(i\right)\psi^{\dagger}\left(j\right)\right]\right\rangle \\
=\theta\left(t_{i}-t_{j}\right)\left\langle \left\{ \psi\left(i\right),\psi^{\dagger}\left(j\right)\right\} \right\rangle \label{eq.G}
\end{multline}
has the following expression 
\begin{multline}
G\left(\mathbf{k},\omega\right)=\frac{1}{\omega-\varepsilon\left(\mathbf{k}\right)+\mathrm{i}\delta}I\left(\mathbf{k}\right)\\
=\sum_{m=1}^{3}\frac{\sigma^{\left(m\right)}\left(\mathbf{k}\right)}{\omega-E_{m}\left(\mathbf{k}\right)+\mathrm{i}\delta}\label{eq:Gk}
\end{multline}
where $E_{m}\left(\mathbf{k}\right)$ are the eigenvalues of the energy
matrix $\varepsilon(\mathbf{k})$ and, as poles of the Green's function,
serve as main excitation bands of the system. $\sigma^{\left(m\right)}\left(\mathbf{k}\right)$
are the spectral density weights per band and can be computed as
\begin{equation}
\sigma_{ab}^{(m)}\left(\mathbf{k}\right)=\sum\limits _{c=1}^{3}\Omega_{am}\left(\mathbf{k}\right)\Omega_{mc}^{-1}\left(\mathbf{k}\right)I_{cb}\left(\mathbf{k}\right)\label{eq:sigmak}
\end{equation}
where the matrix $\Omega\left(\mathbf{k}\right)$ contains the eigenvectors
of $\varepsilon\left(\mathbf{k}\right)$ as columns.

The correlation functions of the fields of the basis $C_{ab}\left(i,j\right)=\langle\psi_{a}\left(i\right)\psi_{b}^{\dagger}\left(j\right)\rangle$
can be easily determined in terms of the Green's function by means
of the spectral theorem and their Fourier transforms have the general
expression 
\begin{align}
C_{ab}\left(\mathbf{k},\omega\right) & =2\pi\sum\limits _{m=1}^{3}C_{ab}^{\left(m\right)}\left(\mathbf{k}\right)\delta\left(\omega-E_{m}\left(\mathbf{k}\right)\right)\label{eq:Ck}\\
C_{ab}^{\left(m\right)}\left(\mathbf{k}\right) & =\left[1-f_{\mathrm{F}}\left(E_{m}\left(\mathbf{k}\right)\right)\right]\sigma_{ab}^{\left(m\right)}\left(\mathbf{k}\right)\label{eq:Ck_m}
\end{align}
where $f_{\mathrm{F}}\left(\omega\right)=\left(\mathrm{e}^{\frac{\omega}{k_{\mathrm{B}}T}}+1\right)^{-1}$
is the Fermi function and $C_{ab}^{\left(m\right)}\left(\mathbf{k}\right)$
is the band component per momentum of the corresponding same-time
correlation function $C_{ab}\left(\mathbf{k}\right)=\sum\limits _{m=1}^{3}C_{ab}^{\left(m\right)}\left(\mathbf{k}\right)$.

\subsection{Normalization $I$ matrix}

In a paramagnetic and homogeneous system, the normalization $I\left(\mathbf{k}\right)$
matrix has the following entries
\begin{align}
I_{11}\left(\mathbf{k}\right) & =I_{11}=1-\frac{n}{2}\label{eq:I11k}\\
I_{12}\left(\mathbf{k}\right) & =0\label{eq:I12k}\\
I_{13}\left(\mathbf{k}\right) & =3C_{\xi c}^{\alpha}+\frac{3}{2}\alpha\left(\mathbf{k}\right)\chi_{s}^{\alpha}\label{eq:I13k}\\
I_{22}\left(\mathbf{k}\right) & =I_{22}=\frac{n}{2}\label{eq:I22k}\\
I_{23}\left(\mathbf{k}\right) & =3C_{\eta c}^{\alpha}-\frac{3}{2}\alpha\left(\mathbf{k}\right)\chi_{s}^{\alpha}\label{eq:I23k}\\
I_{33}\left(\mathbf{k}\right) & =4C_{c_{s}c}^{\alpha}+\frac{3}{2}C_{\eta\eta}+3\alpha\left(\mathbf{k}\right)\left(f_{s}+\frac{1}{4}C_{cc}^{\alpha}\right)\nonumber \\
 & +\frac{3}{2}\beta\left(\mathbf{k}\right)\chi_{s}^{\beta}+\frac{3}{4}\eta\left(\mathbf{k}\right)\chi_{s}^{\eta}\label{eq:I33k}
\end{align}
where $n=\left\langle n\left(i\right)\right\rangle $ is the filling,
$\chi_{s}^{\kappa}=\frac{1}{3}\left\langle n_{k}^{\kappa}\left(i\right)n_{k}\left(i\right)\right\rangle $
is the spin-spin correlation function at distances determined by the
projector $\kappa$ and $f_{s}=\frac{1}{3}\left\langle c^{\dagger}\left(i\right)\cdot\sigma_{k}\cdot c^{\alpha}\left(i\right)n_{k}^{\alpha}\left(i\right)\right\rangle $
is a higher-order (up to three different sites are involved) spin-spin
correlation function. We have also introduced the following definitions,
which is based on those related to the correlation functions of the
fields of the basis (\ref{eq:Ck}): $C_{\phi\varphi}=\left\langle \phi_{\sigma}\left(i\right)\varphi_{\sigma}^{\dagger}\left(i\right)\right\rangle $
and $C_{\phi\varphi}^{\kappa}=\left\langle \phi_{\sigma}^{\kappa}\left(i\right)\varphi_{\sigma}^{\dagger}\left(i\right)\right\rangle $,
where no summation over sigma is intended. $\beta\left(\mathbf{k}\right)$
and $\eta\left(\mathbf{k}\right)$ are the projectors onto the second-nearest-neighbor
sites along the main diagonals and the main axes of the lattice, respectively.

\subsection{$m$-matrix}

In a paramagnetic and homogeneous system, the $m$-matrix has the
following entries
\begin{align}
m_{11}\left(\mathbf{k}\right) & =-\mu I_{11}-4t\left[\Delta+\left(p+I_{11}-I_{22}\right)\alpha\left(\mathbf{k}\right)\right]\label{eq:m11k}\\
m_{12}\left(\mathbf{k}\right) & =4t\left[\Delta+\left(p-I_{22}\right)\alpha\left(\mathbf{k}\right)\right]\label{eq:m12k}\\
m_{13}\left(\mathbf{k}\right) & =-\left(\mu+4t\alpha\left(\mathbf{k}\right)\right)I_{13}\left(\mathbf{k}\right)-4t\alpha\left(\mathbf{k}\right)I_{23}\left(\mathbf{k}\right)\nonumber \\
 & -2tI_{33}\left(\mathbf{k}\right)-4t\alpha\left(\mathbf{k}\right)I_{\bar{\pi}c_{s}}^{\alpha}\label{eq:m13k}\\
m_{22}\left(\mathbf{k}\right) & =\left(U-\mu\right)I_{22}-4t\left[\Delta+p\alpha\left(\mathbf{k}\right)\right]\label{eq:m22k}\\
m_{23}\left(\mathbf{k}\right) & =\left(U-\mu\right)I_{23}\left(\mathbf{k}\right)+2tI_{33}\left(\mathbf{k}\right)+4t\alpha\left(\mathbf{k}\right)I_{\bar{\pi}c_{s}}^{\alpha}\label{eq:m23k}\\
m_{33}\left(\mathbf{k}\right) & =-\mu I_{33}\left(\mathbf{k}\right)+2dtI_{\kappa_{s}c_{s}^{\dagger}}\left(\mathbf{k}\right)+UI_{\eta_{s}c_{s}^{\dagger}}\left(\mathbf{k}\right)\label{eq:m33k}
\end{align}
where $\Delta=C_{\xi\xi}^{\alpha}-C_{\eta\eta}^{\alpha}$ is the difference
between upper and lower intra-Hubbard-subband contributions to the
kinetic energy and $p=\frac{1}{4}\left(\chi_{0}^{\alpha}+3\chi_{s}^{\alpha}\right)-\chi_{p}^{\alpha}$
is a combination of the nearest-neighbor charge-charge $\chi_{0}^{\alpha}=\left\langle n^{\alpha}\left(i\right)n\left(i\right)\right\rangle $,
spin-spin $\chi_{s}^{\alpha}$ and pair-pair $\chi_{p}^{\alpha}=\left\langle \left[c_{\uparrow}\left(i\right)c_{\downarrow}\left(i\right)\right]^{\alpha}c_{\downarrow}^{\dagger}\left(i\right)c_{\uparrow}^{\dagger}\left(i\right)\right\rangle $
correlation functions.

\subsection{Self-consistency and Algebra constraints}

By restricting $I_{\kappa_{s}c_{s}^{\dagger}}\left(\mathbf{k}\right)$
and $I_{\eta_{s}c_{s}^{\dagger}}\left(\mathbf{k}\right)$ to just
the local and the nearest-neighbor terms \cite{Avella_14}, we have
\begin{equation}
m_{33}\left(\mathbf{k}\right)\cong-\mu I_{33}\left(\mathbf{k}\right)+\bar{m}_{33}^{0}+\alpha\left(\mathbf{k}\right)\bar{m}_{33}^{\alpha}\label{eq:m33k_approx}
\end{equation}
and we can use a couple of Algebra constraints\cite{Theory,Avella_11a,Avella_14}
to compute $\bar{m}_{33}^{0}$ and $\bar{m}_{33}^{\alpha}$. $I_{\bar{\pi}c_{s}}^{\alpha}$
can be fixed in the very same manner \cite{Avella_14}. For the sake
of consistency, we also neglect the $\beta\left(\mathbf{k}\right)$
and $\eta\left(\mathbf{k}\right)$ terms in $I_{33}\left(\mathbf{k}\right)$
\cite{Avella_14}. More details can be found in \cite{Avella_14}.
We can recognize the following Algebra Constraints
\begin{align}
C_{\xi\xi} & =1-n+D\label{eq:Cxixi}\\
C_{\eta\eta} & =\frac{n}{2}-D\label{eq:Cetaeta}\\
C_{\xi\eta} & =0\label{eq:Cxieta}\\
C_{\xi c_{s}} & =3C_{\xi c}^{\alpha}\label{eq:Cxics}\\
C_{\eta c_{s}} & =0\label{eq:Cetacs}
\end{align}
where $D=\left\langle n_{\uparrow}\left(i\right)n_{\downarrow}\left(i\right)\right\rangle $
is the double occupancy. These relations lead to the following very
relevant ones 
\begin{align}
n & =2\left(1-C_{\xi\xi}-C_{\eta\eta}\right)\label{eq:n}\\
D & =1-C_{\xi\xi}-2C_{\eta\eta}\label{eq:D}
\end{align}

On the other hand, we can compute $\chi_{0}^{\alpha}$, $\chi_{s}^{\alpha}$,
$\chi_{p}^{\alpha}$ and $f_{s}$ by operatorial projection, which
is equivalent to the well-established one-loop approximation \cite{Theory,Avella_11a}
for same-time correlations functions 
\begin{align}
\chi_{0}^{\alpha} & \approx n^{2}-2\frac{I_{11}\left(C_{c\eta}^{\alpha}\right)^{2}+I_{22}\left(C_{c\xi}^{\alpha}\right)^{2}}{C_{\eta\eta}}\label{eq:chia0}\\
\chi_{s}^{\alpha} & \approx-2\frac{I_{11}\left(C_{c\eta}^{\alpha}\right)^{2}+I_{22}\left(C_{c\xi}^{\alpha}\right)^{2}}{2I_{11}I_{22}-C_{\eta\eta}}\label{eq:chias}\\
\chi_{p}^{\alpha} & \approx\frac{C_{c\xi}^{\alpha}C_{\eta c}^{\alpha}}{C_{\eta\eta}}\label{eq:chiap}\\
f_{s} & \approx-\frac{1}{2}C_{c\xi}^{\alpha}-\frac{3}{4}\chi_{s}^{\alpha}\left(\frac{C_{c\xi}^{\alpha}}{I_{11}}-\frac{C_{c\eta}^{\alpha}}{I_{22}}\right)\nonumber \\
 & -2\frac{C_{c\xi}^{\alpha}}{I_{11}}\left(C_{c\xi}^{\alpha^{2}}-\frac{1}{4}C_{c\xi}\right)-2\frac{C_{c\eta}^{\alpha}}{I_{22}}\left(C_{c\eta}^{\alpha^{2}}-\frac{1}{4}C_{c\eta}\right)\label{eq:fs}
\end{align}
Summarizing, we can fix the unknowns $I_{\bar{\pi}c_{s}}^{\alpha}$,
$\bar{m}_{33}^{0}$, $\bar{m}_{33}^{\alpha}$, $\mu$, $\chi_{0}^{\alpha}$,
$\chi_{s}^{\alpha}$, $\chi_{p}^{\alpha}$ and $ $$f_{s}$ through
the set of equations (\ref{eq:Cxieta}), (\ref{eq:Cxics}), (\ref{eq:Cetacs}),
(\ref{eq:n}), (\ref{eq:chia0}), (\ref{eq:chias}), (\ref{eq:chiap})
and (\ref{eq:fs}).

\section{Results\label{sec:Results}}

\subsection{Double occupancy and chemical potential: solution assesment\label{sec:Results:D_mu}}

\begin{figure}[tp]
\begin{centering}
\includegraphics[width=7cm]{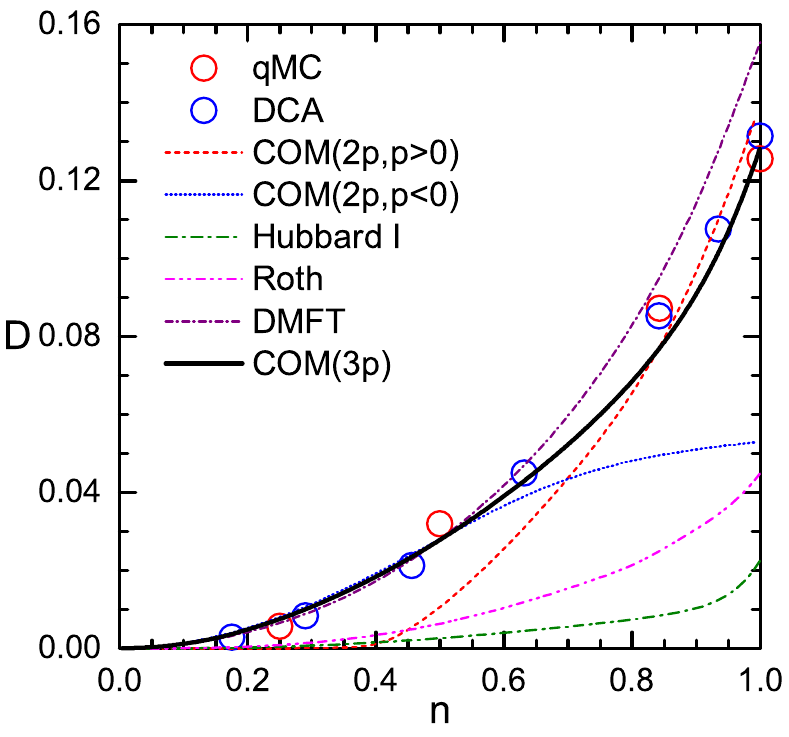}
\par\end{centering}

\begin{centering}
\includegraphics[width=7cm]{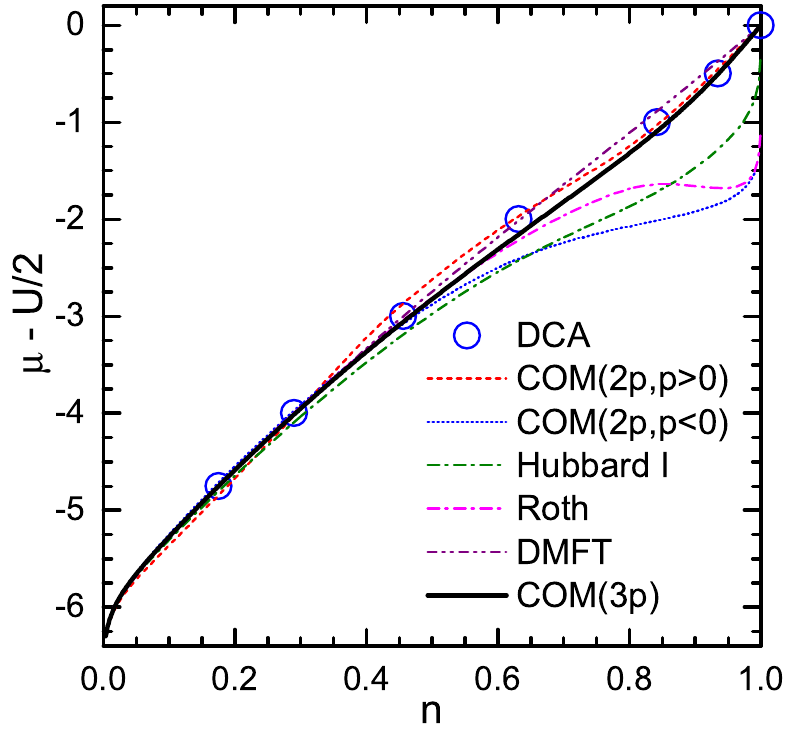}
\par\end{centering}

\protect\caption{Double occupancy $D$ (left) and scaled chemical potential $\mu-\nicefrac{U}{2}$
(right) as functions of the filling $n$ for $U=4$ and $T=\nicefrac{1}{6}$
for COM(3p) (black lines), COM(2p,$p>0$) (dashed red line), COM(2p,$p<0$)
(dotted blue line), Hubbard I (dot-dashed green line), Roth (dot-dot-dashed
magenta line) and DMFT \cite{Capone} (dash-dotted purple line). Analytical
results are compared with $12\times12$-site qMC \cite{Moreo_90}
and $2$-site DCA \cite{Sangiovanni} numerical data (red and blue
hollow circles, respectively).\label{fig:1_2}}
\end{figure}

In Fig.\ \ref{fig:1_2}, we report the behavior of the double occupancy
$D$ (left panel) and of the scaled chemical potential $\mu-\nicefrac{U}{2}$
(right panel) as functions of the filling $n$ for $U=4$ and $T=\nicefrac{1}{6}$.
It is evident the very good agreement in the whole range of filling
$n$ between COM(3p) and the $12\times12$-site qMC \cite{Moreo_90}
and $2$-site DCA \cite{Sangiovanni} numerical data. The double occupancy
$D$ features a very elaborated behavior presenting a continuos, but
well defined, change of slope on approaching half filling. COM(3p)
correctly catches this feature, while all other presented solutions
do not manage to achieve the same level of agreement over the whole
range of filling. Hubbard I and Roth solutions report values of the
$D$ extremely far from the numerical ones and always much smaller
than these latter, showing a tendency to an excess of correlations
present in such solutions. DMFT \cite{Capone} performs extremely
well, with respect to numerical data, at low-intermediate values of
filling, but at intermediate-high ones features values of $D$ larger
than the numerical ones. This is a clear evidence of a lack of correlations
for this value of $U$. COM(2p, $p<0$) performs really very well
too at low-intermediate values of filling, but on increasing $U$
it shows an excess of correlations close to half filling (it is actually
insulating for any finite value of $U$ at half filling). In COM(2p,
$p>0$), it is evident a complete suppression of $D$ at low values
of the filling as well as a small, but visible, discrepancy in the
slope close to half filling. COM(3p) evidently has the capability
to correctly interpolate between the two COM(2p) solutions sticking
to COM(2p, $p<0$) at low-intermediate values of filling and even
improving on COM(2p, $p>0$) at intermediate-high values of filling.
The DCA data for the chemical potential show a concavity in proximity
of half filling that is correctly caught by COM(3p) and COM(2p, $p>0$)
and not by COM(2p, $p<0$), Hubbard I and Roth solutions. Roth solution
actually reports a rather evident region of thermodynamic instability,
$\frac{d\mu}{dn}<0$, close to half filling. As a matter of fact,
$U=4$ induces already quite strong electronic correlations: the chemical
potential gets ready to open a gap for higher values of $U$ and $n=1$.
COM(2p, $p<0$), Hubbard I and Roth solutions place themselves always
on the strongly correlated side and report values of $\mu$ quite
far from the numerical ones: their particle counting - actual effective
filling - is definitely far from the exact one. DMFT \cite{Capone}
solution does not catch the correct concavity again showing a lack
of correlations for this value of $U$, but it features values of
$\mu$ very close to the numerical ones in the whole range of filling
$n$ although not so close as COM(3p) ones in proximity of half-filling,
which is the most interesting region.

\subsection{Quasi-particle energy dispersions: solution characterization\label{sec:Results:Ek}}

\begin{figure}[tp]
\begin{centering}
\includegraphics[width=8cm]{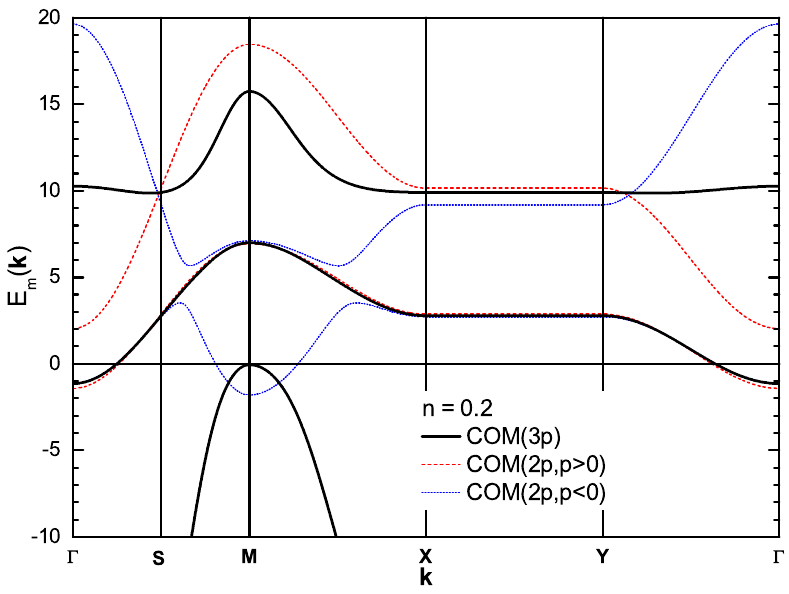}
\par\end{centering}

\begin{centering}
\includegraphics[width=8cm]{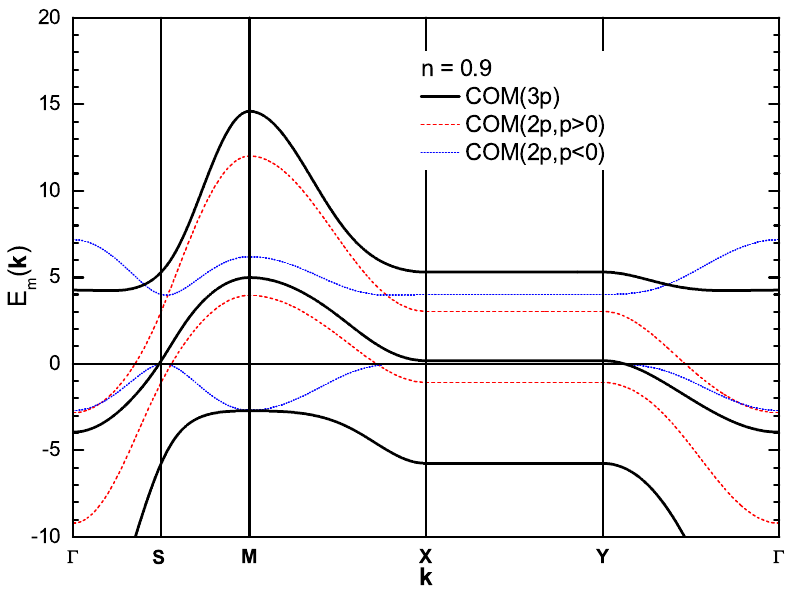}
\par\end{centering}

\protect\caption{Energy bands $E_{m}\left(\mathbf{k}\right)$ along the principal directions
of the first Brillouin zone ($\Gamma=(0,0)$ $\to$ $S=(\nicefrac{\pi}{2},\nicefrac{\pi}{2})$
$\to$ $M=(\pi,\pi)$ $\to$ $X=(\pi,0)$ $\to$ $Y=(0,\pi)$ $\to$
$\Gamma=(0,0)$) at $T=\nicefrac{1}{6}$, $U=4$ and two different
values of the filling $n=0.2$ (left) and $n=0.9$ (right) for COM(3p)
(black line), COM(2p,$p>0$) (red line) and COM(2p,$p<0$) (blue line).\label{fig:3_4}}
\end{figure}

In Fig.\ \ref{fig:3_4}, we report the energy bands $E_{m}\left(\mathbf{k}\right)$
along the principal directions of the first Brillouin zone ($\Gamma=(0,0)$
$\to$ $S=(\nicefrac{\pi}{2},\nicefrac{\pi}{2})$ $\to$ $M=(\pi,\pi)$
$\to$ $X=(\pi,0)$ $\to$ $Y=(0,\pi)$ $\to$ $\Gamma=(0,0)$) at
$T=\nicefrac{1}{6}$, $U=4$ and two different values of the filling
$n=0.2$ (left panel) and $n=0.9$ (right panel). At $n=0.2$, it
is evident that the occupied bands are almost identical across all
reported COM solutions. Actually, COM(3p) is characterized by a small,
but finite, occupation of its LHB, besides the occupation of its central
band (CB), which is the band coinciding with the COM(2p) LHBs. This
can be understood in terms of the proximity of COM(3p) LHB to the
chemical potential at the $M$ point. LHB is the only occupied band
in COM(2p,$p<0$) at all finite values of $U$. At $n=0.9$, the occupied
region in energy-momentum space across the three COM solutions is
instead quite different, although some similarities can still be found.
In particular, as regards the regions close to the chemical potential
at the $\Gamma$ point and along the main anti-diagonal (the $X-Y$
line). COM(3p) CB, which was the main actor at low fillings, tends
to systematically lose occupation in favor of the LHB. Close to half
filling, this latter eventually exceeds the former in occupation and
collects more and more of it on increasing $U$ while the CB depletes
on approaching the metal-insulator transition. As regards COM(2p,$p>0$)
instead, UHB plays a minor role all the way up to the metal-insulator
transition. It collects a small fraction of the electronic occupation
and only above a certain intermediate value of the filling. It is
evident that COM(3p) CB is still almost pinned to the chemical potential
along the main anti-diagonal (the $X-Y$ line); the van Hove singularity
lies little below the Fermi level. Accordingly, changing the filling
in this region of low doping (from $n=0.85$ to $n=1$) has mainly
the effect to induce a transfer of spectral weight between the bands
and between their components in terms of fields of the basis, as one
would expect in a strongly correlated regime, rather than shifting
the chemical potential more or less rigidly within the bands, as it
could be expected at small fillings and weak interactions. It is also
evident that the LHB has still a minor role with respect to the CB,
which collects the vast majority of the occupied states. It is worth
noting that the spin-spin correlations are already present, but not
yet so strong to determine the reduction of the bandwidth in the energy-momentum
space region shared by CB and LHB. It is worth noticing that COM(3p)
bands are quite close to COM(2p,$p<0$) ones.

\subsection{Momentum resolved quantities\label{sec:Results:nk_Dk_chiask}}

\begin{figure*}[tp]
\begin{centering}
\begin{tabular}{ccc}
\includegraphics[width=0.33\textwidth]{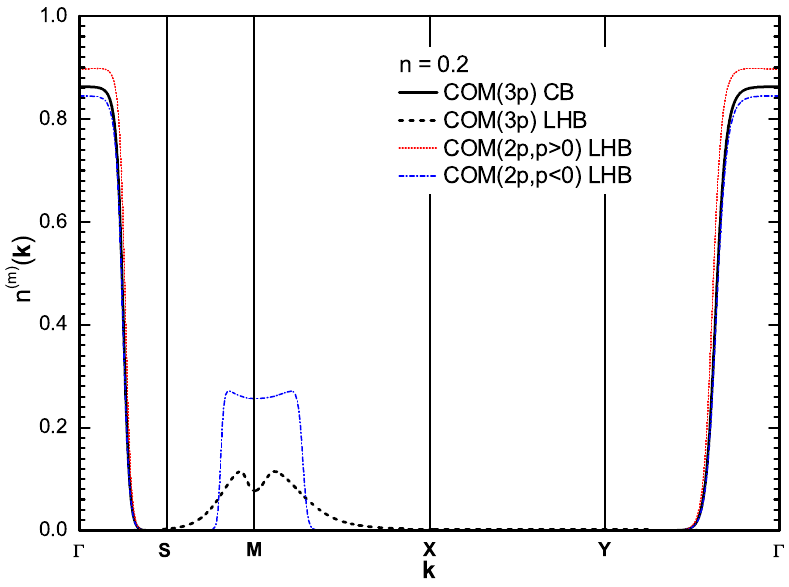} & \includegraphics[width=0.33\textwidth]{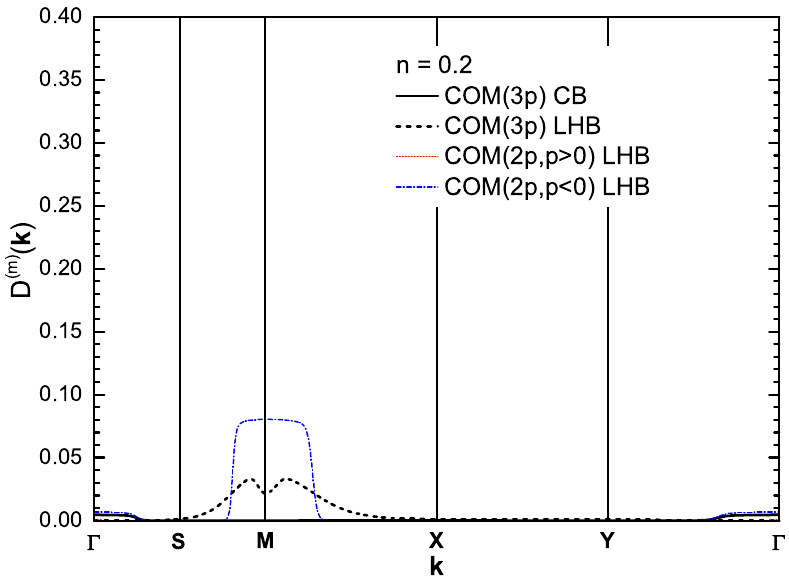} & \includegraphics[width=0.33\textwidth]{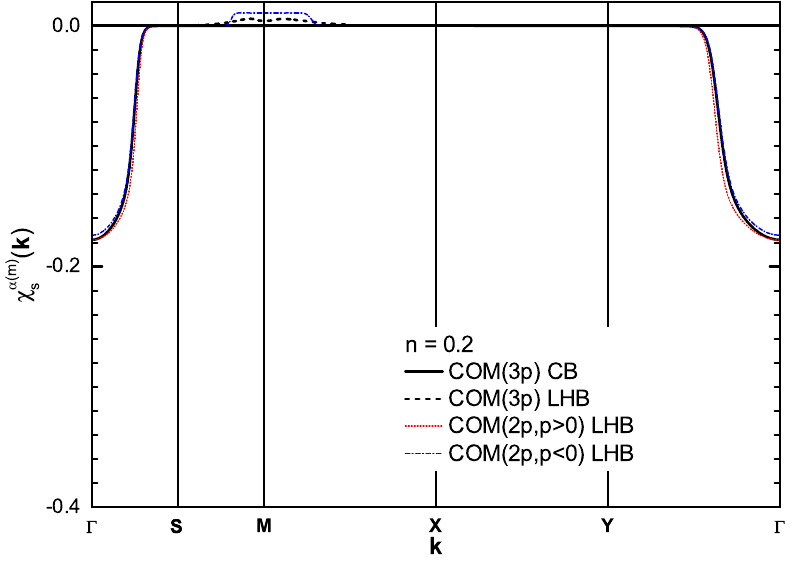}\tabularnewline
\includegraphics[width=0.33\textwidth]{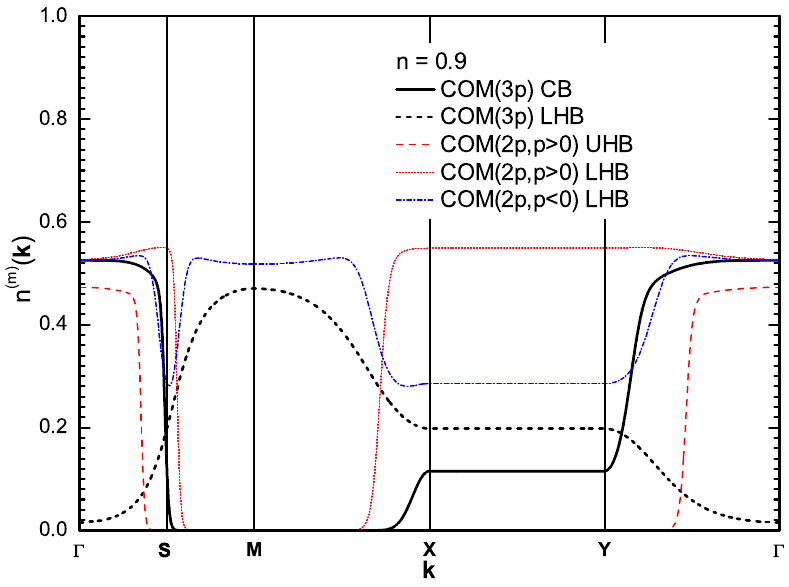} & \includegraphics[width=0.33\textwidth]{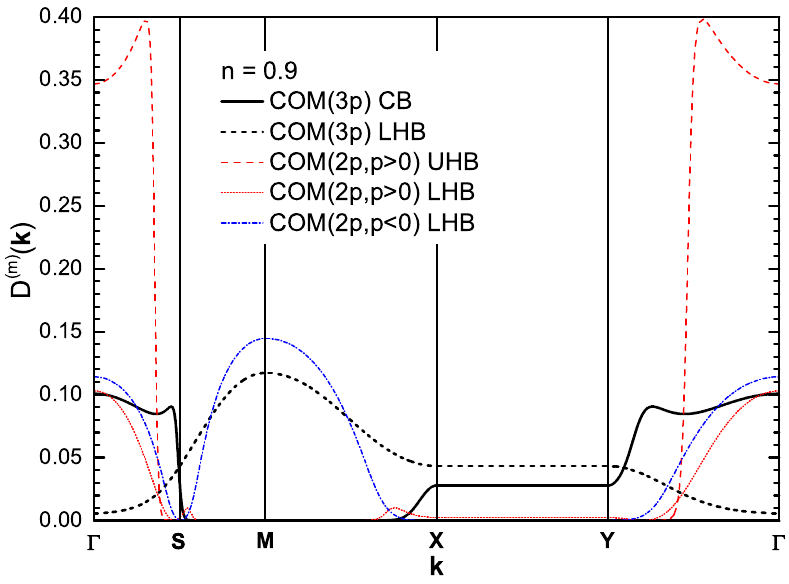} & \includegraphics[width=0.33\textwidth]{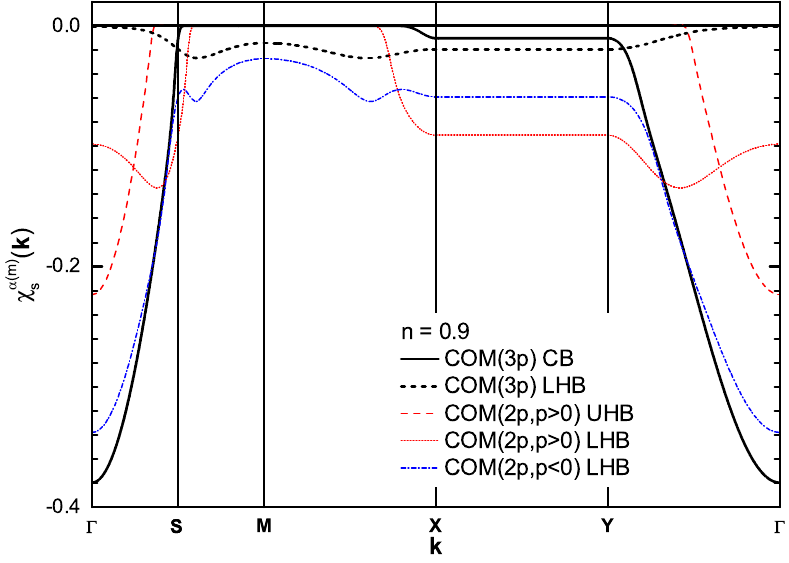}\tabularnewline
\end{tabular}
\par\end{centering}

\protect\caption{Momentum-distribution function per band and spin $n^{\left(m\right)}\left(\mathbf{k}\right)$
(left), double-occupancy components per band $D^{\left(m\right)}\left(\mathbf{k}\right)$
(center) and nearest-neighbor spin correlation function components
per band $\chi_{s}^{\alpha\left(m\right)}\left(\mathbf{k}\right)$
(right) along the principal directions of the first Brillouin zone
($\Gamma=(0,0)$ $\to$ $S=(\nicefrac{\pi}{2},\nicefrac{\pi}{2})$
$\to$ $M=(\pi,\pi)$ $\to$ $X=(\pi,0)$ $\to$ $Y=(0,\pi)$ $\to$
$\Gamma=(0,0)$) at $T=\nicefrac{1}{6}$, $U=4$ and two different
values of the filling $n=0.2$ (top) and $n=0.9$ (bottom) for COM(3p)
(solid and short-dashed black line), COM(2p,$p>0$) (dashed and short-dotted
red line) and COM(2p,$p<0$) (short-dash-dotted blue line).\label{fig:5_10}}
\end{figure*}

Given the decomposition of the momentum-dependent correlation functions
per band reported in Eqs.\ \ref{eq:Ck} and \ref{eq:Ck_m}, it is
possible to define a similar decomposition for any quantity that can
be expressed in terms of correlation functions of the chosen operatorial
basis, that is for any quantity computable within the reported approximation.
In particular, we have the following expressions for the filling $n$,
the double occupancy $D$ and the nearest-neighbor spin correlation
function $\chi_{s}^{\alpha}$
\begin{align}
n & =2\sum\limits _{m=1}^{3}\frac{1}{N}\sum_{\mathbf{k}}n^{\left(m\right)}\left(\mathbf{k}\right)
\end{align}
\begin{align}
D & =\sum\limits _{m=1}^{3}\frac{1}{N}\sum_{\mathbf{k}}D^{\left(m\right)}\left(\mathbf{k}\right)
\end{align}
\begin{align}
\chi_{s}^{\alpha} & =\sum\limits _{m=1}^{3}\frac{1}{N}\sum_{\mathbf{k}}\chi_{s}^{\alpha\left(m\right)}\left(\mathbf{k}\right)
\end{align}
where
\begin{align}
n^{\left(m\right)}\left(\mathbf{k}\right) & =\frac{1}{2}f_{\mathrm{F}}\left(E_{m}\left(\mathbf{k}\right)\right)\sigma_{cc}^{\left(m\right)}\left(\mathbf{k}\right)\\
D^{\left(m\right)}\left(\mathbf{k}\right) & =f_{\mathrm{F}}\left(E_{m}\left(\mathbf{k}\right)\right)\sigma_{22}^{\left(m\right)}\left(\mathbf{k}\right)\\
\chi_{s}^{\alpha\left(m\right)}\left(\mathbf{k}\right) & =\frac{2\alpha\left(\mathbf{k}\right)f_{\mathrm{F}}\left(E_{m}\left(\mathbf{k}\right)\right)}{2I_{11}I_{22}-C_{\eta\eta}}\times\nonumber \\
 & \times\left(I_{11}C_{c\eta}^{\alpha}\sigma_{2c}^{\left(m\right)}\left(\mathbf{k}\right)+I_{22}C_{c\xi}^{\alpha}\sigma_{1c}^{\left(m\right)}\left(\mathbf{k}\right)\right)
\end{align}

In Fig.\ \ref{fig:5_10}, we report the momentum-distribution function
per band and spin $n^{\left(m\right)}\left(\mathbf{k}\right)$ (left
column), the double-occupancy components per band $D^{\left(m\right)}\left(\mathbf{k}\right)$
(center column) and the nearest-neighbor spin correlation function
components per band $\chi_{s}^{\alpha\left(m\right)}\left(\mathbf{k}\right)$
(right column) along the principal directions of the first Brillouin
zone ($\Gamma=(0,0)$ $\to$ $S=(\nicefrac{\pi}{2},\nicefrac{\pi}{2})$
$\to$ $M=(\pi,\pi)$ $\to$ $X=(\pi,0)$ $\to$ $Y=(0,\pi)$ $\to$
$\Gamma=(0,0)$) at $T=\nicefrac{1}{6}$, $U=4$ and two different
values of the filling $n=0.2$ (top row) and $n=0.9$ (bottom row).
Components from not reported bands are zero or definitely negligible.

At $n=0.2$, $n^{\left(m\right)}\left(\mathbf{k}\right)=\frac{1}{2}\mathcal{F}_{\mathbf{k}}\left[\left\langle c^{\dagger}(\mathbf{i})\cdot c(\mathbf{j}\right\rangle )\right]_{m}$
shows that reported COM bands have a similar and quite \emph{ordinary}
occupations except for the region close to $M$ that is occupied only
for COM(3p) and COM(2p,$p<0$) LHBs. This is the result of the peculiar
shape of such bands (see Fig.\ \ref{fig:3_4}) that closely recalls
the bending driven by antiferromagnetic fluctuations and the simultaneous
occupation of $\Gamma$ and $M$ points. What is really surprising
is the fact that the major, almost the only, contribution to the double
occupancy ($D^{\left(m\right)}\left(\mathbf{k}\right)=\frac{1}{2}\mathcal{F}_{\mathbf{k}}\left[\left\langle \eta^{\dagger}(\mathbf{i})\cdot\eta(\mathbf{j}\right\rangle )\right]_{m}$)
comes just from the this last region in momentum (close to $M$ point).
This is really counterintuitive as one expects those branches of energy-momentum
dispersion to have such a shape because of the strong antiferromagnetic
fluctuations and to be the main seat of electrons contributing to
single occupation and, therefore, with well-formed spin momenta. As
a matter of fact, the contribution to $D^{\left(m\right)}\left(\mathbf{k}\right)$
close to $M$ should be correctly read as the main (actually the only)
contribution to the kinetic energy coming from the $\eta$ Hubbard
operators, that is from those electrons moving between double occupied
sites. In fact, this situation well explains the almost identical
value of the double occupancy for COM(3p) and COM(2p,$p<0$) at this
filling, where the motion between doubly occupied sites is allowed,
as well as the almost negligible value for COM(2p,$p>0$), where the
motion between doubly occupied sites is definitely negligible as it
is confined to the UHB, which is empty for this value of the filling.
The decomposition of $\chi_{s}^{\alpha\left(m\right)}\left(\mathbf{k}\right)$
shows the expected negative contribution by the electrons close to
$\Gamma$ point and an almost negligible positive contribution by
those close to $M$ point. At such low value of the filling, we can
expect very weak spin correlations and the contribution close to $\Gamma$
point is not so large as well as the whole momentum dependence very
little structured.

At $n=0.9$, $n^{\left(m\right)}\left(\mathbf{k}\right)$ shows how
much COM bands differ in occupation between the three reported solutions
for a value of the filling where quite intense correlations are expected.
COM(3p) features an occupation of the CB close to $\Gamma$ point
quite reduced with respect to that sported at $n=0.2$, even if we
take into account that it is now spanning a quite larger region in
momentum and the main anti-diagonal (the $X-Y$ line) is somewhat
filled too. This can be explained by noting that the LHB, which at
$n=0.2$ was filled only close to $M$ point, features now a quite
relevant occupation spanning all over the first Brillouin zone and,
in particular, at the $M$ point and along the main anti-diagonal
(the $X-Y$ line). Accordingly, we expect the physics of COM(3p) solution
to be determined by both bands at the same time and on almost equal
footing. Overall, COM(2p,$p<0$) occupation is very similar to the
COM(3p) one although concentrated in the only occupied band, the LHB.
As a matter of fact, COM(2p,$p<0$) LHB seems to interpolate between
the LHB and the CB of COM(3p) showing once more the very strict connections
between these two solutions. COM(2p,$p>0$) features instead similar
occupations for the two bands except for the extension towards the
main anti-diagonal (the $X-Y$ line) of the almost completely filled
LHB. The region in momentum close to $M$ remains anyhow empty marking
the greatest difference to the other two solutions. Coming to $D^{\left(m\right)}\left(\mathbf{k}\right)$,
COM(3p) solution features again a complementary presence between CB
and LHB with the exception of the main anti-diagonal (the $X-Y$ line)
where the more marked difference reported for $n^{\left(m\right)}\left(\mathbf{k}\right)$
is greatly reduced. Bare looking at the values of $D^{\left(m\right)}\left(\mathbf{k}\right)$
reported along these principal direction, COM(2p,$p<0$) should have
an overall value of $D$ quite similar to that of COM(3p), but this
is quite not right and can be explained by looking at the region in
momentum along the main anti-diagonal (the $X-Y$ line) and close
to it (even along the main diagonal: the $\Gamma-M$ line). Although,
COM(2p,$p<0$) LHB is lying over the chemical potential in this region
of momentum, the double occupancy contribution is definitely negligible
marking a huge difference to the CB of COM(3p) that occupies the same
region in momentum-frequency space. This is one of the net improvements
of the COM(3p) solution over COM(2p,$p<0$) one; an improvement that
reflects in many other physical quantities. COM(2p,$p>0$) contributions
come both (for LHB and UHB) from the region in momentum close to $\Gamma$
clearly showing that although the overall value of $D$ is quite similar,
clearly by accident, between COM(3p) and COM(2p,$p>0$), its physical
origin, in terms of quasi-particle contributions and of their momentum-frequency
dispersions, is very different and explains the quite different behavior
in terms of slope of $D$ as a function of the filling $n$. Let us
come to $\chi_{s}^{\alpha\left(m\right)}\left(\mathbf{k}\right)$.
As regards COM(3p), the CB brings a much larger contribution, with
respect to $n=0.2$, that extends along the main anti-diagonal (the
$X-Y$ line). LHB contribution is much smaller, but definitely larger
than at $n=0.2$ and negative. Accordingly, it is just the CB, which
originates from the third basic field describing the nearest-neighbor
antiferromagnetic fluctuations, to bring the larger contribution as
one would have expected. Therefore, having such a field in the basis
results as one of the main ingredients in order to get such a good
performance in comparing this solution with numerical ones \cite{Avella_14}.
COM(2p,$p<0$) contributions to $\chi_{s}^{\alpha\left(m\right)}\left(\mathbf{k}\right)$
all come from the only occupied band, the LHB, and once more seem
to mime the overall behavior of COM(3p). COM(2p,$p>0$) has a completely
different behavior. In particular, the contribution of the LHB close
to the $\Gamma$ point is quite difficult to understand: instead of
increasing in absolute value towards the $\Gamma$ point, it decreases
leading to the presence of a maximum absolute value for a value of
momentum that coincide with the Fermi surface of the related COM(2p,$p>0$)
UHB.

\section{Summary\label{sec:Summary}}

In this manuscript, we have first recollected the main analytical
expressions defining a recently proposed, within the framework of
the Composite Operator Method\cite{Theory,Avella_11a}, three-pole
solution for the two-dimensional Hubbard model \cite{Avella_14}.
Together with the two Hubbard fields, well describing the physics
at the energy scale of $U$, the presence of a third field, embedding
the strong antiferromagnetic fluctuations, has enormously boosted
the performance of COM(3p) solution with respect to COM(2p) ones.
The extremely positive comparison with the data obtained by different
numerical methods for momentum-integrated quantities (e.g. local properties)
as functions of all model parameters (filling, on-site Coulomb repulsion
and temperature) as well as for the energy bands of the system \cite{Avella_14}
makes this solution extremely interesting to be analyzed further.
Here, we have reported a summary of the behavior of the basic local
quantities - the double occupancy and the chemical potential -, together
with the quasi-particle energy dispersions definitely necessary to
guide the subsequent analysis, which is the main focus of the present
manuscript: the study of the momentum-resolved components of filling
(i.e. the momentum distribution function), double occupancy and nearest-neighbor
spin correlation function. The analysis has been extended to COM(2p)
solutions that have been used as primary reference as in the main
paper \cite{Avella_14}. Analyzing the momentum-resolved quantities,
it emerges very clearly the role played by the third field with respect
to the two Hubbard ones in determining the behavior of many relevant
quantities and allowing to get the extremely good comparison with
numerical results. In particular, the proximity between COM(3p) and
COM(2p,$p<0$) solutions is further reinforced and better understood
giving a guideline to further improve and, possibly, optimize the
application of the COM to the Hubbard model with the choice of a fourth
field solving the few remaining issues with COM(3p)\cite{Avella_14}.
\begin{acknowledgments}
The author wishes to thank Gerardo Sica for many insightful discussions.
\end{acknowledgments}


\end{document}